\documentstyle[aas2pp4,epsf]{article}		% approximate journal style

\received{}
\accepted{}
\lefthead{Church et al.}
\righthead{Simple Photoelectric Absorption during Dipping} 

\begin{document}

\title{Simple Photoelectric Absorption during Dipping 
in the ASCA Observation of XB\thinspace 1916-053}
\author{M. J. CHURCH\altaffilmark{1,2}, 
T. DOTANI\altaffilmark{2}, M. BA\L UCI\'NSKA-CHURCH\altaffilmark{1,2}, 
K. MITSUDA\altaffilmark{2},
\hbox {T. TAKAHASHI\altaffilmark{2},} H. INOUE\altaffilmark{2} 
AND K. YOSHIDA\altaffilmark{3} }

\altaffiltext{1}{School of Physics and Space Research, University of
Birmingham, Edgbaston, Birmingham B15 2TT, UK}
\altaffiltext{2}{Institute of Space and Astronautical Science, Yoshinodai
3-1-1, Sagamihara, Kanagawa 229, JAPAN}
\altaffiltext{3}{Faculty of Engineering, Kanagawa University, 3-27-1
 Rokkakubashi, Kanagagawa-ku, Yokohama 221, JAPAN}

% The abstract environment prints out the receipt and acceptance dates
% if they are relevant for the journal style.  For the aasms style, they
% will print out as horizontal rules for the editorial staff to type
% on, so long as the author does not include \received and \accepted
% commands.  This should not be done, since \received and \accepted dates
% are not known to the author.

\begin{abstract}
We report results of analysis of the ASCA observation of the Low Mass
X-ray Binary dipping source XB\thinspace
1916-053 made on 1993, May 2nd, during which dipping was very deep
such that in the deepest parts of dips, 
the X-ray intensity in the band 0.5 - 12.0 keV fell to zero,
demonstrating that all emission components were completely removed.
The best-fit orbital period of the binary system determined from the X-ray
data was found to be $\rm {3005 \pm 10}$~s.  The high quality ASCA 
data allowed spectral evolution in dipping to be systematically
investigated by spectral analysis in intensity bands covering the full
range of dipping from intensities close to zero to non-dip values.
We have shown that the spectra can be well fitted by the same 
two-component model previously used to give good explanations of
the very different dip sources X\thinspace 1755-338 and X\thinspace
1624-490, consisting of point-source blackbody emission from the neutron
star plus extended Comptonised emission probably from the accretion disk
corona. In the case of XB\thinspace 1916-053 we show that all levels of
dipping can be fitted using $\rm {kT_{bb}}$ = $\rm {2.14\pm 0.28}$ keV 
and power law photon index = $\rm {2.42\pm 0.21}$ 
which are the best-fit values for non-dip data,
together with the corresponding non-dip normalisations. Dipping is
shown to be due to large increases of column density for the point-like
blackbody, combined with the extended power law component being
progressively covered by absorber until in the deepest parts of dips,
the partial covering fraction approaches unity. This approach differs radically
from the ``absorbed plus unabsorbed'' approach previously used in
spectral modelling of XB\thinspace 1916-053 and similar sources, in which
the normalisation of the unabsorbed component is allowed to decrease
markedly in dipping, behavior generally attributed to the
effects of electron scattering. Thus we have shown that spectral
evolution in XB\thinspace 1916-053 can be explained simply in terms of
photoelectric absorption without the need for substantial electron scattering.
This explanation is supported by calculation of the relative importance 
of photoelectric absorption and electron scattering in the absorbing 
region which shows that little electron scattering is expected in the ASCA 
energy band.

\end{abstract}

% The different journals have different requirements for keywords.  The
% keywords.apj file, found on aas.org in the pubs/aastex-misc directory, 
% contains a list of keywords used with the ApJ and Letters.  These are 
% usually assigned by the editor, but authors may include them in their 
% manuscripts if they wish. 


\section{Introduction}
XB\thinspace 1916-053 is an important member of the class of $\sim $ 10
Low Mass X-ray Binary (LMXB) sources that exhibit decreases in X-ray
intensity at the orbital period, generally accepted as being due to
absorption in the bulge in the outer accretion disk where the flow from
the companion impacts on the outer disk (White and Swank, 1982). 
XB\thinspace 1916-053 is unusual in several respects: it has the shortest
orbital period of the dipping sources, 50 min (Walter et al. 1982).
It has a depth of dipping that is highly variable which at times reaches 
100\%. The source is also notable because of the difference between the X-ray 
period and the optical period of $\sim $ 1\% (Grindlay et al. 1988).

\medskip
\noindent
Three Exosat observations were made and the results of Smale et al. (1988) 
showed that the average extent of dipping changed markedly between the 
observations as shown by comparing the light curves folded on the X-ray 
period for the 3 observations.
\begin{figure*}[t]
\leavevmode\epsffile{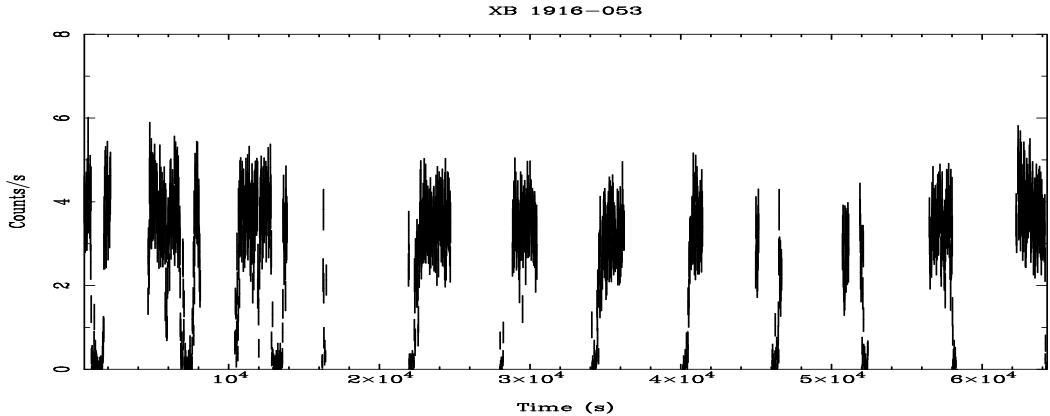}
\caption{ASCA GIS2 light curve for the complete 18~hr observation in the
energy band 0.5 - 12.0 keV with 16~s binning. \label{fig1}}
\end{figure*}
Spectral analysis showed that non-dip data for all 3 observations
was best-fitted by a simple power law model with photon index $\Gamma $
close to 1.80. Dip spectra were selected in intensity bands from all 3 
observations, for which the best fit was a model consisting of two power
laws, each with the index fixed at the above value.
The column density of the one component could be held at the
quiescent value, while that of the other component increased strongly
in dipping; ie there was an absorbed and an unabsorbed term. 
For the unabsorbed component, the normalisation decreased strongly
in dipping. This ``absorbed plus unabsorbed'' approach has been used 
for several of the dip sources. It has been applied to the sources: 
XBT\thinspace 0748-676 (Parmar et al. 1986), X\thinspace 1254-690 
(Courvoisier et al. 1986) and  X\thinspace 1624-490 (Jones and Watson,
1989).
It is clear that the parameters of the source emitting regions cannot change 
coincidentally with dipping and the strong decrease in normalisation of the 
unabsorbed component has been attributed to the effects of electron scattering
in the absorber (Parmar et al 1986; Smale et al. 1988); 
ie there is a decrease in flux from the source 
due to scattering which does not reveal itself as low energy absorption. 
More recently, GINGA data on XB\thinspace 1916-053 has also been fitted by
the absorbed plus unabsorbed approach (Smale et al. 1992; 
Yoshida et al., 1995). Yoshida et al. showed 
that the variation in normalisation can be reproduced
as absorption by cold matter if electron scattering in the absorber is
taken into account.

\medskip 
\noindent
XB\thinspace 1916-053 shows increases in hardness during dip ingress. 
It was originally expected that all of the dipping sources should show 
hardening during dipping, due to photoelectric absorption of the X-ray
spectrum in the absorbing bulge which preferentially removes the lower
energy X-rays.
However, the dip sources do not in general follow this expectation; 
some sources show a hardening, but some show a marked softening in dipping, 
eg X\thinspace 1624-490 (Church and Balucinska-Church, 1995) which is 
totally unexpected on simple physical models, ie with a single emission
component. Furthermore X\thinspace 1755-338 has dipping which is independent 
of energy in the band 1 - 10 keV (White et al. 1984).

\medskip
\noindent
Several types of spectral model have been used in fitting the dipping sources.
The Exosat spectra of several dipping sources were fitted by a
Comptonisation model (White et al. 1988) represented by a single component
absorbed cut-off power law. Other workers have used  two-component
models, notably Mitsuda and co-workers (see for example, Mitsuda et
al. 1984). Apart from this difference in approach, individual dipping 
sources have generally been fitted by different spectral models. 
In particular,
the absorbed plus unabsorbed approach has involved different spectral
forms when applied to different sources; ie power law plus power law,
cut-off power law plus cut-off power law, etc. More recently,
Church and Balucinska-Church (1993, 1995) have proposed a 
two-component or ``complex continuum'' model which has 
been able to explain the softening in dipping in X\thinspace 1624-490
and the energy-independence of dipping in X\thinspace 1755-338
using the same model. In this model, X-ray emission
originates as blackbody radiation from the boundary layer at the surface
of the neutron star, plus power law emission representing Comptonisation in
an accretion disk corona (valid at energies much lower than the
Comptonisation break). In the above sources, the model shows that dipping
is primarily due to absorption of the point-source blackbody emission
with comparatively little absorption of the extended power law emission.
Whether there is  softening, hardening or energy-independence depends
mostly on the blackbody temperature. In X\thinspace 1624-490, this is higher
than in X\thinspace 1755-338 ($\rm {kT_{bb}}$ = 1.39 keV compared with 
0.88 keV), so that removal of the blackbody leaves the
residual spectrum softer than in non-dip emission. 

\vfill \eject
%\medskip
\noindent
The main aim of the present work was to test the complex continuum model
in the case of XB\thinspace 1916-053 using high quality ASCA data, and
to test whether this physical model can offer an alternative to the
absorbed plus unabsorbed approach.

%\placefigure{1}

\section{Results}
The 18~hr observation of XB\thinspace 1916-053 was made on 1993, May 2nd. with
the satellite ASCA (Tanaka et al. 1994). The quality of the data was very good, 
and we present here results primarily for analysis of the GIS2 and GIS3
detector data (see Ohashi et al. 1996). The data were screened 
to remove regions of South Atlantic
Anomaly passage, to restrict elevation of the source above the limb of
the Earth to more than 5$^\circ $, particle cut-off rigidity to more than 6,
angular deviations of the telescope pointing 
to less than 0.6 arcmin, and the radiation belt parameter to less than 500.
Rise-time discrimination was applied to remove detector particle
background. Data were pre-selected from the image to remove the region of the
calibration source and outer ring in the image, and source data were 
taken from a 6 arcmin radius circle centered on the source.
Various tests were made on the best way of subtracting background, and
eventually we found that the most reliable background subtraction
could be obtained by taking background data from a circular region
of radius 8 arcmin offset from the center of the image by the same amount 
as the source region (about 2 arcmin) diametrically opposite to the source
containing no visible point sources.
This region provided a sufficiently large count for background subtraction
which is important, with a count rate of 0.13 c/s.

\medskip
\noindent
The total observation of XB\thinspace 1916-053 spanned about 22 orbital
cycles. The good data inbetween SAA passage and Earth occultation included
parts of about 10 dips. Much of this dip data was very fragmented; however
during the first 5 orbital cycles observed, complete data were obtained
for 2 orbital cycles. In addition complete data were
obtained from this part of the observation for one more dip. Burst data
were contained in the observation, but did not remain in GIS data after
screening and so did not affect our spectral analysis.

\begin{figure}[h]
\leavevmode\epsffile{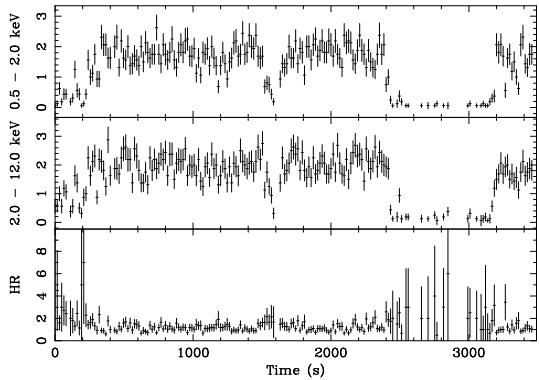}
\caption{GIS2 light curves for the 3rd section of data in Fig. 1 in the 
energy bands 2.0 - 12.0 keV and 0.5 - 2.0 keV and the hardness ratio formed 
by dividing these. \label{fig2}}
\end{figure}

\subsection {Light Curves and Hardness Ratio}  
Light curves were extracted in various energy bands and it was found that
during part of every dip, the count rate
fell to zero in the total band of the GIS detectors 0.5 - 12.0 keV. 
The background-subtracted light curve of the GIS2 data is shown in Fig. 1, 
from which the good coverage of 3 dips in the early part of the observation 
can be seen, and the relatively poor definition of later dips. Interdips
between the main dips can also be seen.

\medskip
\noindent
Hardness ratio was obtained by dividing the c/s in the band 2.0 - 12.0 keV 
by the
c/s in the band 0.5 - 2.0 keV, and the light curves in these bands and the 
plot of hardness ratio against time are shown in Fig. 2 for one
orbital cycle including non-dip, interdip and dip data. 
Increases in hardness ratio at dip ingress and egress and in interdips
can be seen; during deep dipping when all count rates were very
small the hardness ratio is not well-determined. 
The depth of dipping defined as the maximum percentage decrease
in count rate was investigated as a function of energy by plotting light
curves in several energy bands. In the lower bands 0.5 - 2.0 keV, 2.0 -
4.0 keV and 4.0 - 6.0 keV, dipping was 100\%; in the higher band 6.0 -
12.0 keV, the depth possibly decreased to $\rm {95\pm 5}$\%. Thus dipping 
was remarkably deep
at all energies implying an absorber of high column density completely
covering all emission regions. In the interdips dipping was not 100\%.

\begin{figure}[h]
\leavevmode\epsffile{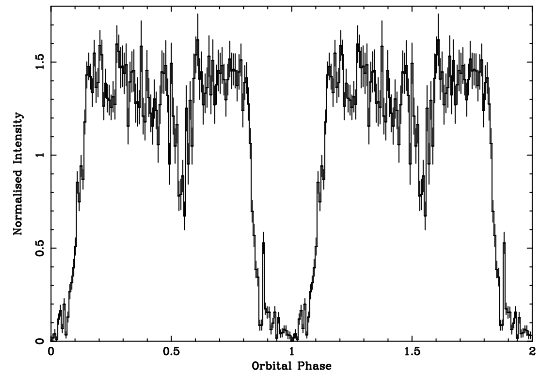}
\caption{ASCA GIS2 data folded on the 3005~s best-fit orbital
period. \label{fig3}}
\end{figure} 

%\placefigure{2}

\subsection {Light Curves Folding Analysis}
For determining the best-fit X-ray orbital period, we used fast mode SIS0 data 
for the whole observation with time resolution of 1~s. The data were
folded on trial periods spanning the range 2900~s to 3100~s to find the
most probable period, from which a best value 
of $\rm {3005 \pm 10}$ s was obtained. In Fig. 3 we show 
the GIS2 light curve from the first 5 orbital periods in the
observation folded on this period (to avoid severe distortion of the folded
light curve by adding in data not containing interdips, for example). 
It is difficult to be certain
of the errors attached to this period. The errors of $\rm {\pm 10}$~s
represent the range of possible periods around the most likely value
in the plot of $\chi^2$ {\it versus} period produced by the period searching 
program, consistent with the scatter of the $\chi ^2$ points. However, 
the period is being determined using dip data of variable length, depth  
and quality, and the errors are inevitably larger than in fitting, for 
example, a coherent oscillation. By folding the data on periods outside 
the range of the errors above (within $\rm {3005 \pm 20}$~s) it was found 
that folded light curves having shapes consistent with the raw light curves
could be produced, so that the real errors may be somewhat larger than
those quoted. Previous values for the X-ray period were determined, for example, 
from {\it Exosat} data to be $\rm {3015 \pm 17}$~s, and from GINGA data to be 
$\rm {3005 \pm 6.6}$~s (Smale et al. 1992). From Fig. 3, it can be seen that the
main dip lasts 33\% of the orbital cycle implying an absorbing region
that subtends an angle of $\rm {120^{\circ}}$ at the neutron star. The
consequence of this is that the absorbing region must be very extended
azimuthally and may therefore also be extended in height above the disk.
Thus it might be expected that the depth of dipping can reach 100\% 
as observed, since all emission regions can be covered by the
absorber during dips, ie not just point source emission from the neutron
star, but also extended emission regions. This will be relevant to the
spectral modelling discussed below.
%\placefigure{3}

\subsection {Spectral Evolution in Dipping}
The very large depth of dipping in XB\thinspace 1916-053 during the ASCA
observation, the complete coverage of several dips and the high quality
of the data, have allowed detailed investigation of spectral evolution in
dipping by dividing the data into intensity bins. GIS2 data were divided
into bands corresponding to non-dip data: 4.0 - 5.0 c/s and various levels
of dipping at 3.0 - 4.0 c/s, 2.0 - 3.0 c/s, 1.0 - 2.0 c/s and 0.0 - 1.0 c/s
and spectra produced. 
The corresponding GIS3 data were selected by using the time filters
produced when GIS2 data were selected into intensity bands. Because of
differences between the detectors, this GIS3 data formed bands somewhat 
different in intensity from the original GIS2 bands. GIS2 and GIS3 spectra
were added, systematic errors of 2\% added to each channel and the data
regrouped to a minimum of 100 counts per bin. Channels below 0.7 keV and 
above 10.0 keV were ignored.
Response functions and effective areas were used in the spectral
fitting as appropriate to the regions from which the data were extracted,
and the GIS2 and GIS3 effective area files combined.
Background spectra from GIS2 and GIS3 were combined appropriately.
Non-dip and dip data were fitted by a number of spectral models, 
beginning with various simple models such as an absorbed power law, 
absorbed blackbody and absorbed bremsstrahlung.
The absorbed power law model gave a formally acceptable fit to all intensity
bands except the lowest intensity band for which the reduced $\chi ^2_r$
was 2.8; however the power law photon index varied between 1.76 in non-dip
and 0.61 in deep dipping which is clearly not acceptable physically.
The absorbed blackbody could not fit non-dip data with a $\chi ^2_r$
of 9.9, and the absorbed bremsstrahlung model had $\chi^2_r$ values
in the range 1 - 4.7 for various intensity bands and kT values varying
between 9.4 and 200 keV. Thus it is clear that simple models cannot be used. 
Consequently, we next attempted to fit the kind of two-component model
used for the sources X\thinspace 1755-33 and
X\thinspace 1624-490 (Church \& Balucinska-Church 1993, 1995),
consisting of a blackbody component associated with the
neutron star and a power law component associated 
probably with the accretion disk corona. Each component was
given its own absorption term and the model may be represented as:
$\rm {AB_1*BB + AB_2*PL}$. The two sources above could be fitted well by
this model with $\rm {kT_{bb}}$ and $\Gamma $ and the normalisations
of the 2 components fixed at the non-dip values; dipping was due primarily
to incresases of column density for the blackbody.
In the case of XB\thinspace 1916-053, the dip data could not be fitted
with the normalisations held constant, with $\chi^2_r$ values for deep
dip data as large as 19.3. This was related to the appearance of
a double-humped shape to the dip spectra, ie the apparent existence
of an unabsorbed component at lower energies, or soft excess. In the
absorbed plus unabsorbed approach, the low energy hump is the
unabsorbed component which persists into dipping even though the
higher energy hump is strongly absorbed. As we were unable
to fit the two-component model with normalisations fixed at the
non-dip values, we next fitted the same model with free normalisations.
This model gave an acceptable fit to all intensity levels. 
The normalisation of the
power law component decreased by a large factor and the blackbody
column density increased by a large amount in dipping.
Thus this is a form of the absorbed plus unabsorbed
approach which represents a halfway
stage between the normal absorbed and unabsorbed approach and our final 
best-fit modelling below.

\medskip
\noindent
It is an integral part of the complex continuum model that the
source emission parameters cannot change during
dipping, and so the normalisations should be held constant.
However during dipping the count rate falls to zero, and the model must
reproduce this. Thus in this particular source the appropriate form of the
complex continuum model to use must allow for all emission regions to
be covered by the absorber, and so we applied the model in the following
form: $\rm {AB_1}$*BB + AG*[$\rm {AB_2}$*f + (1-f)]*PL 
where AG represents Galactic absorption
and f is a partial covering fraction. The blackbody absorption term, of
course, also includes Galactic absorption; however we have investigated the
best values of column density of the two components as discussed
below, and this was done by varying the AG term.
With the above spectral model, a good fit was obtained to the 
non-dip data giving values for the blackbody temperature 
$\rm {kT_{bb}}$ = 2.14$\pm 0.28$ and power law index 
$\Gamma $ = 2.42$\pm 0.21$. 
With $\rm {kT_{bb}}$ and $\Gamma $ and the corresponding normalisations held
fixed at the best-fit non-dip values, very good fits could be obtained
to all intensity levels of dip data, with $\chi^2_r$ always less than 0.84.
Extensive testing was carried out to determine whether interdips should be included
with main dip data. It was found that differences in spectral fitting
parameters between dip data and interdip data at the same intensity were 
generally within 90\% confidence errors, although there was some tendency 
for blackbody column densities to be higher in dips. To obtain 
better statistics, the final analysis was carried out including all dips 
and interdips. 
The best $\rm {N_H}$ values for the blackbody and power law components
for the non-dip spectrum were determined by fixing the AG term in the 
spectral model at various values, and determining $\chi^2_r$ and the
total column density of each component.  

\begin{figure}[h]
\leavevmode\epsffile{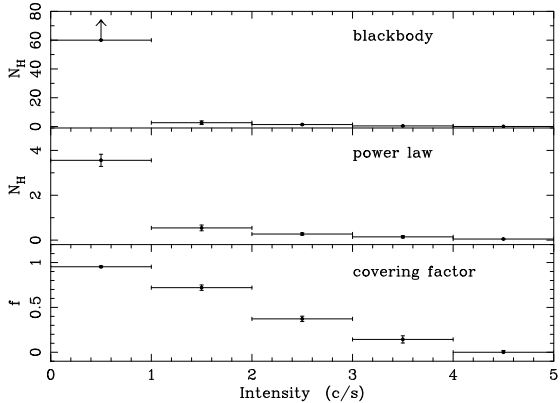}
\caption{Best-fit spectral fiting results : variation of blackbody column 
density, power law column density and power law partial covering fraction 
{\it f} with intensity. $\rm {N_H}$ is in units of $\rm {10^{23}}$ H atom
$\rm {cm^{-2}}$.
\label{fig4}}
\end{figure} 

%\placefigure{4}

\noindent
When the AG term was fixed at $\rm {2\cdot
10^{21}}$ H atom $\rm {cm ^{-2}}$ (the Stark et al. (1992) value), 
the blackbody column density became
very large at $\rm {60\cdot 10^{21}}$ H atom $\rm {cm^{-2}}$
with $\chi^2_r$ equal to 9.0. As the
AG term was increased, $\chi^2_r$ decreased 
to a value of $\sim $ 0.78 when the blackbody and power law column
densities were equal with a value of $\rm {4.75 \cdot 10^{21}}$ H atom $\rm
{cm^{-2}}$.
This value is in excess of the Stark et al. value implying a 
degree of intrinsic absorption in the source during this observation. 
Plots of the spectral fitting results as a function of X-ray
intensity are shown in Fig. 4, and parameter values are given in Table 1.
\placetable{1}
%\placefigure{5}

\begin{table}
\caption{Best fit spectral fitting results.
\label{1}}
\begin{tabular} {llrll}
$\rm {I^a}$ & $\rm {N_H(BB)^b}$ &$\rm {N_H(PL)^b}$ & f & $\chi^2_r$ \\
\tableline
4 - 5  & 0.48$\pm $0.18   & 0.48$\pm $0.18 & $\rm {0.00^{+0.02}_{-0.0}}$ & 0.78 \\
3 - 4  & 4.3$\pm $0.6   & 1.5$\pm $0.5 & 0.14$\pm $0.04 & 0.84 \\
2 - 3  & 13.8$\pm $2.7   & 2.7$\pm $0.5 & 0.37$\pm $0.03 & 0.84 \\
1 - 2  & 26.7$\pm $12.6   & 5.4$\pm $1.3 & 0.72$\pm $0.03 & 0.67 \\
0 - 1  & $>$600           & 35.6$\pm $2.7  & 0.95$\pm $0.01 & 0.49 \\
\tableline
\end{tabular}

$\rm {^a}$ intensity in c/s; $\rm {^b}$ in units of $\rm {10^{22}}$ 
H atom $\rm {cm^{-2}}$
\end{table}

\begin{figure}[h]
\leavevmode\epsffile{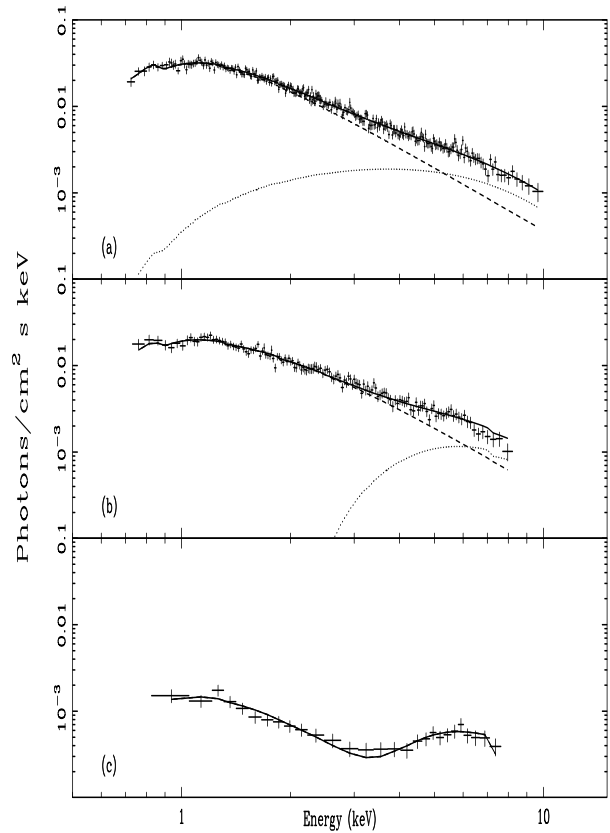}
\caption{Contributions of the blackbody (dots) and the power law (dashes)
to the net photon spectrum for (a) the non-dip spectrum, (b) the
intensity band 2.0 - 3.0 c/s and (c) the intensity band 0.0 - 1.0 c/s.
Note that in (c) the blackbody is totally absorbed.
\label{fig5}}
\end{figure} 

\medskip
\noindent
Fig. 5 shows source spectra for non-dip, intermediate dip and deep dip
data, together with the 2 model components, to illustrate how the
spectral components of the complex continuum model
can successfully model the dipping. 
It can be seen that there is little evidence for spectral line
features in the non-dip data. Because of its
relatively high temperature ($\rm {kT_{bb}}$ = 2.14 keV), the blackbody
peaks at $\sim $ 6.5 keV, ie in the higher energy part of the ASCA
spectrum. As this component is point-like it is immediately covered by
absorber and $\rm {N_H}$ increases by about a factor of 25 
as the intensity falls from 4.0 - 5.0 c/s to 2.0 - 3.0 c/s. 
The power law, however, starts with a partial covering fraction of
zero and at first suffers little absorption. Thus, as dipping develops,
the power law remains as a low energy peak while at higher energies the
blackbody is strongly absorbed. In deeper
dipping, $\rm {N_H}$ for the blackbody increases to at least $\rm
{6000\cdot 10^{21}}$ H atom $\rm {cm^{-2}}$, completely removing this
component. The partial covering fraction 
reaches 0.95, $\rm {N_H}$ for the power law also becomes high at 
$\rm {350\cdot 10^{21}}$ H atom $\rm{cm^{-2}}$, and the power law
is split between the low energy hump which is the 5\% not covered,
and the higher energy hump which is covered, but not totally absorbed.
Finally, the fluxes of both components become zero.
Thus there are no changes in normalisation, and the dipping
can be explained simply in terms of photoelectric absorption of the 
point source blackbody emission from the neutron star combined with
absorption of the extended power law component by a relatively large
absorbing bulge such that the extended component is progressively
covered by the absorber. In this model, the ``absorbed'' part is the
blackbody plus the covered part of the power law, and the ``unabsorbed''
part is the uncovered power law.

\section{Discussion}
We have demonstrated that the two-component model
can give a good description of spectral evolution in XB\thinspace 1916-053.
In this model,
emission originates as point-source blackbody emission from the neutron
star plus extended power law emission probably from the accretion disk
corona. In the case of 
XB\thinspace 1916-053, the source intensity in the band 1.0 - 10.0 
keV often actually becomes zero in the deepest parts of dips. 
In X\thinspace 1755-338 and X\thinspace 1624-490, dipping was not 100\%
and spectral evolution during dips was explained by the 2-component model
with dipping being primarily due to absorption of the blackbody
(Church and Balucinska-Church 1993, 1995).
In the case of XB\thinspace 1916-053, it was necessary to 
allow the spectral modelling to take account of the fact that the power law
component must be totally removed in deepest dipping. The modelling we
performed showed that it was not sufficient to use a spectral form AB*BB
+ AB*PL; the data could not be fitted by this model with normalisations
fixed.  It was necessary to allow the extended power law component in our
model to be progressively covered by the absorber; extended emission 
would not be covered essentially instantaneously as is the point-source 
component. With the inclusion of the partial covering term, the two-component 
model provides very good fits to the spectra.

\medskip
\noindent

\medskip
\noindent
The fluxes of the blackbody and power law components in the non-dip
data in the energy range 1 - 10 keV are $\rm {0.98\cdot 10^{-10}}$ erg
$\rm {cm^{-2}\;s^{-1}}$ and $\rm {1.83\cdot 10^{-10}}$ erg $\rm
{cm^{-2}\;s^{-1}}$ respectively, so that the blackbody contributes
34\% to the total energy flux in this band. In X\thinspace 1755-338 and
X\thinspace 1624-490, in which the two-component model showed that dipping was
due primarily to absorption of the blackbody, the spectral
evolution in dipping is determined by $\rm {kT_{bb}}$; in X\thinspace
1624-490 $\rm {kT_{bb}}$ was 1.39 keV, 
the blackbody peaking at $\sim $4.5 keV, ie it was relatively
hard such that the residual power law spectrum when the blackbody was 
absorbed was softer. XB\thinspace 1916-053 has even higher $\rm {kT_{bb}}$
of 2.14 keV; however this does not determine the spectral 
evolution during dips since both components are absorbed. The low energy
cut-off of the spectrum is determined by the power law component, and
the hardening observed at dip ingress is clearly simply due to 
absorption of the low energy part of the spectrum.

\medskip
\noindent
Perhaps the most interesting question is whether
we expect electron scattering to be important in the absorbing region
producing the dips. In other dip sources in which the absorbed plus
unabsorbed approach was not used, electron scattering was not thought
to be important. Electron scattering may take place in XB\thinspace
1916-053 between the source regions and the absorbing bulge in the outer
disk, but this will occur in both non-dip and dip cases and so is not
relevant. In the absorbing region, we can determine the state of
ionization by estimating the ionization parameter $\xi $ as follows, using
the column density of the point source blackbody component as a probe 
of density along a track through the absorbing bulge as dipping develops.
As dipping develops, parts of the absorber at different radial
positions from the center of the absorber will contribute to
attenuation of the incident radiation, having different lengths
along the line of sight.
We can write $\xi = {L\epsilon/N_H r}$ where L is the
luminosity, r is radial distance from the source and the thickness of
the absorber along the line of sight is a fraction $\epsilon $ of the
accretion disk radius, assumed to fill approximately the Roche lobe of 
the neutron star.
From our best-fit modelling, the unabsorbed flux of the source 
in the band 1 - 20 keV is 
$\rm {4.1\cdot 10^{-10}}$ erg $\rm {cm^{-2}\;s^{-1}}$,
and the corresponding luminosity is $\rm {3.4\cdot 10^{36}}$ erg
$\rm{s^{-1}}$, using a distance of 8.4 kpc (Smale et al. 1988).
As dipping commences, in the intensity band 3.0 - 4.0 c/s, the
blackbody has column density $\rm {4.3\cdot 10^{22}}$ H atom $\rm
{cm^{-2}}$ and estimating $\epsilon $ as 0.03, we
find that $\xi $ = 70 erg cm $\rm {s^{-1}}$. 
For this value, ionization will not be complete. In deepest dipping,
$\rm {N_H}$ rises to at least $\rm {6\cdot 10^{24}}$ H atom $\rm
{cm^{-2}}$, and the thickness of absorber along the line of sight
will be maximum, so if we take $\epsilon $ $\sim $ 0.3 we get $\xi $
= 5, such that some elements only can be, at most, singly ionized.

\medskip \noindent
In both cases described above where ionization is not complete,
the relative importance of photoelectric absorption and electron scattering
is given by the ratio ${\rm {N_H\cdot \sigma _{PE}/ N_e\cdot
\sigma_{T}}}$, where $\sigma_{PE}$ is the total photoelectric absorption
cross section, $\sigma_{T}$ is the Thomson scattering cross section, 
and $\rm {N_e}$ is the electron column density. This follows from the
dependences of the processes on $\rm {exp -(N_H\sigma_{PE})}$
and $\rm {exp-(N_e\sigma_{T})}$.

For a completely ionized plasma of a medium with Solar abundances,
it can be calculated that the electron density $\rm {n_e}$ is related to
the ion density $\rm {n_i}$ via $\rm {n_e \; \simeq \; 1.2\cdot n_i}$,
since elements other than H contribite more than 1 electron, but with
small abundances. Thus in the above cases $\rm {N_H \; \simeq \; N_e}$
and the above ratio 
is dominated by the ratio of cross sections. At 1 keV, 
$\sigma _{PE}/\sigma _T$  $\simeq $ 500, at 4 keV, $\sigma
_{PE}/\sigma _T$ $\simeq $ 10, and at 10 keV, $\sigma _{PE}/\sigma _T$
$\simeq $ 2. Thus photoelectric absorption strongly dominates over
electron scattering throughout most of the ASCA band. We should also
consider how this depends on $\rm {N_H}$. 
As the column density increases in deep dipping corresponding to the
central regions of the absorber, at $\rm {N_H}$
 = $\rm {1.5\cdot 10^{24}}$ H atom $\rm {cm^{-2}}$, the
optical depth for electron scattering becomes unity. However the product
$\rm {N_H \sigma_{PE}}$ will be very much greater than 1; 
ie the probability of photoelectric
absorption is still much higher than that of electron scattering. If
incident photons are allowed to be simultaneously absorbed and scattered,
in deep dipping the factor exp-$\rm {(N_e \sigma_T)}$ 
implies that an appreciable
fraction of the radiation incident upon the absorber could be scattered.
The average column density of $\rm {35.6\cdot 10^{22}}$ H atom $\rm
{cm^{-2}}$ from spectral fitting
implies a loss of 20\% by scattering, and it could be argued that the
spectral model we have applied should be modified for deep dipping.
However if there is a density gradient in the absorber, with $\rm {N_H}$ =
$\rm {10^{22}}$ - $\rm {10^{23}}$ H atom $\rm {cm^{-2}}$ in the outer
layers, the optical depth for absorption will be $>$ 1, but
for scattering $<$ 1. Thus photons will be preferentially removed
before they reach the higher density central regions where scattering
would play a part. At 10 keV, this preferential absorption will be
much weaker, so that in deepest dipping, electron scattering of the
high energy photons will take place. The blackbody emission is by this
stage of dipping already highly absorbed leaving flux only in the range
just below 10 keV. Thus our value of $\rm {N_H}$ for the blackbody in deepest
dipping of $>$ $\rm { 600\cdot 10^{22}}$ H atom $\rm {cm^{-2}}$ 
may be an overestimate as some decrease
of normalisation may be appropriate for this term, which would imply a
smaller $\rm {N_H}$ value.

\medskip
\noindent
Thus, in summary, we have shown that the physical model used by Church
and Balucinska-Church to explain the dipping sources X\thinspace 1755-338
and X\thinspace 1624-490 also provides a good explanation for
XB\thinspace 1916-053. In all of these cases, the model can be expressed
as point source blackbody emission plus extended Comptonised emission
which is modified by partial covering. For the first two sources the
partial covering fraction f is small, but for XB\thinspace 1916-053
f becomes large in dipping.
Consequently it is possible to explain the dipping
in XB\thinspace 1916-053 purely in terms of photoelectric absorption in a
bulge in the outer accretion disk. There is no need for there to
be substantial electron scattering, and we have shown that in the ASCA 
band little scattering is, in fact, expected.
The explanation of 3 very different members of
the dipping class by the two-component model makes it increasingly likely
that it will be able to explain all members of the class.

\end{document}